\documentclass{aa} 
 
\usepackage{graphicx}
\usepackage{multirow}
\usepackage{hyperref}
\usepackage{amsmath}
\usepackage{nccmath}
\usepackage[flushleft]{threeparttable}
\usepackage{txfonts}
\usepackage{natbib}

\bibpunct{(}{)}{;}{a}{}{,}

\begin{document}

   \title{Complex organic molecules in protoplanetary disks: X-ray photodesorption from methanol-containing ices \\\vspace{0.5cm} \Large Part I - Pure methanol ices} 

   \author{     R. Basalg\`{e}te, \inst{1}    
            R. Dupuy, \inst{1}
            G. F\'{e}raud, \inst{1}
            C. Romanzin, \inst{2}
            L. Philippe, \inst{1}
            X. Michaut, \inst{1}
            J. Michoud, \inst{1}
            L. Amiaud, \inst{3}
            A. Lafosse, \inst{3}
            J.-H. Fillion,  \inst{1}
            M. Bertin \inst{1}                               
                  }

     \offprints{romain.basalgete@obspm.fr}

         \institute {$^1$ Sorbonne Universit\'{e}, Observatoire de Paris, PSL University, CNRS, LERMA, F-75014, Paris, France \\
         $^2$ Univ Paris Saclay, CNRS UMR 8000, ICP, F-91405, Orsay, France \\
         $^3$ Univ Paris Saclay, CNRS, ISMO, F-91405, Orsay, France}
         
     
     \date{Received 14 October 2020 ; Accepted 5 January 2021}

 \abstract{Astrophysical observations show complex organic molecules (COMs) in the gas phase of protoplanetary disks. X-rays emitted from the central young stellar object (YSO) that irradiate interstellar ices in the disk, followed by the ejection of molecules in the gas phase, are a possible route to explain the abundances observed in the cold regions. This process, known as X-ray photodesorption, needs to be quantified for methanol-containing ices. This paper I focuses on the case of X-ray photodesorption from pure methanol ices.}{We aim at experimentally measuring X-ray photodesorption yields (in molecule desorbed per incident photon, displayed as molecule/photon for more simplicity) of methanol and its photo-products from pure CH$_3$OH ices, and to shed light on the mechanisms responsible for the desorption process.}{We irradiated methanol ices at 15 K with X-rays in the 525 - 570 eV range from the SEXTANTS beam line of the SOLEIL synchrotron facility. The release of species in the gas phase was monitored by quadrupole mass spectrometry, and photodesorption yields were derived.}{Under our experimental conditions, the CH$_3$OH X-ray photodesorption yield from pure methanol ice is $\sim 10^{-2}$ molecule/photon at 564 eV. Photo-products such as CH$_4$, H$_2$CO, H$_2$O, CO$_2$ , and CO also desorb at increasing efficiency. X-ray photodesorption of larger COMs, which can be attributed to either ethanol, dimethyl ether, and/or formic acid, is also detected. The physical mechanisms at play are discussed and must likely involve the thermalization of Auger electrons in the ice, thus indicating that its composition plays an important role. Finally, we provide desorption yields applicable to protoplanetary disk environments for astrochemical models.}{The X-rays are shown to be a potential candidate to explain gas-phase abundances of methanol in disks. However, more relevant desorption yields derived from experiments on mixed ices are mandatory to properly support the role played by X-rays in nonthermal desorption of methanol (see paper II).} 

   
             
   \titlerunning{X-ray photodesorption from methanol-containing ices}  
   \authorrunning{Basalg\`ete et al.}  
 
   \maketitle

\section{Introduction} 

Methanol (CH$_3$OH) is a main complex organic molecule (COM) observed in the interstellar medium (ISM) and is considered a precursor for the formation of larger COMs such as, potentially, amino-acids \citep{garrod_complex_2008, elsila_mechanisms_2007}. Its detection in the gas phase of the ISM is often used as a reference to probe other gaseous COMs \citep{bergner_organic_2019}, but its weak emission lines make the observations difficult (e.g., \citealt{carney_upper_2019}). In protoplanetary disks, methanol, formaldehyde (H$_2$CO) and formic acid (HCOOH) have been detected in the gas phase around different young stellar objects (YSOs). In the Class II TW Hydrae protoplanetary disk, a peak column density of $\sim3$-$6 \times 10^{12}$ cm$^{-2}$, suggested to be located at the CO snowline, was derived for methanol in the gas phase \citep{walsh_first_2016}, and formic acid was also detected with a disk-averaged column density of $\sim2$-$4 \times 10^{12}$ cm$^{-2}$ \citep{favre_first_2018}. In the disk around the young Herbig Ae star HD 163296 \citep{carney_upper_2019}, an upper limit on the disk-averaged column density of methanol was found to be $\sim5 \times 10^{11}$ cm$^{-2}$ and a CH$_3$OH/H$_2$CO abundance ratio <0.24 was estimated, compared to a ratio of 1.27 for the TW Hya disk. In the protoplanetary disk around DG Tau, a T Tauri star, disk-height-averaged column densities of H$_2$CO and CH$_3$OH were estimated at $\sim0.3-4 \times 10^{14}$ cm$^{-2}$ and $< 0.04-0.7 \times 10^{14}$ cm$^{-2}$ , respectively, with a CH$_3$OH/H$_2$CO abundance ratio < 1 and a H$_2$CO ring located beyond the CO snow line \citep{podio_organic_2019}. In the circumstellar envelope of the post-asymptotic giant branch (AGB) object HD 101584 \citep{olofsson_first_2017}, H$_2$CO, H$_2^{13}$CO, and CH$_3$OH were also observed.
\\\\
Methanol is the only COM that has been detected as a significant constituent of the icy dust grains that can be found at several stages of star formation, with an estimated abundance between $\sim$ 1\% and $\sim$ 25\% relative to H$_2$O \citep{taban_stringent_2003, gibb_interstellar_2004, boogert_observations_2015}. Beyond the CO snow line, at T $\sim$ 17 K \citep{oberg_competition_2005}, in the coldest outer regions of protoplanetary disks, most COMs are believed to be formed directly onto dust grains, but they can also originate from competitive gas-phase reaction channels depending on the local physical conditions. This is in contrast with methanol, which is generally thought to be entirely formed on the icy grain mantles by successive hydrogenation of CO in the upper CO-rich phase of interstellar ices \citep{watanabe_efficient_2002}, although this is still debated \citep{dartois_non-thermal_2019}. In these cold regions, nonthermal processes should then be invoked to explain the presence of gaseous methanol. 
\\\\
It is expected that photons and/or cosmic rays coming from various sources could trigger the ejection of methanol from the icy dust grains into the gas phase and participate in the overall gas-to-ice balance of these cold regions. For the specific case of photons, this mechanism is known as photodesorption. Several experimental studies have been conducted to quantitatively constrain these processes in order to explain astrophysical observations, especially for methanol-containing ices. In one of these experiments, heavy ion $^{136}$Xe$^{23+}$ irradiations were performed on methanol in pure ice and embedded in a water-ice matrix \citep{dartois_non-thermal_2019}. A sputtering yield of methanol close to that of the main water-ice matrix \citep{dartois_cosmic_2018}, which is $\sim 10^4$ sputtered molecule per incident ion, was estimated for each studied ice. When it was embedded in a CO$_2$ ice, \citet{dartois_non-thermal_2020} found that this sputtering yield is about six times higher. Experimental studies of UV photodesorption in the 7 - 10.5 eV range were first conducted for pure methanol ice at 20 K by \citet{oberg_formation_2009}. More recent experiments by \citet{bertin_uv_2016} in the 7-14 eV range suggested an efficiency for methanol UV photodesorption from pure methanol ice of $\sim10^{-5}$ molecule/photon. This desorption was found to be below the detection threshold ($<10^{-6}$ molecule/photon) when methanol was mixed with CO ice for a wide range of dilution factors (from 1 in 4 to 1 in 50; \citealt{bertin_uv_2016}). Accordingly, \citet{cruz-diaz_negligible_2016} derived an upper limit of $3 \times 10^{-5}$ molecule/photon for the UV photodesorption yield of methanol from pure methanol ice from 8 K to 130 K (in the 6.88-10.9 eV range). In addition to these previous mechanisms, chemical desorption, which is the desorption induced by exothermic reactions, is a possible route for explaining the gas-phase enrichment in the ISM \citep{cazaux_dust_2016, minissale_hydrogenation_2016, ligterink_methanol_2018}. However, chemical desorption of methanol by H addition onto CO, H$_2$CO, and CH$_3$OH ices was not detected (upper limit $<$5\%; \citealt{minissale_hydrogenation_2016}), and more generally, chemical desorption of several other molecules appears to be very low when experiments are made on water substrates \citep{minissale_dust_2016}. 
\\\\
Several observational studies have shown that YSOs (Class I, Class II, and Class III) are X-ray emitters in the range of $\sim$0.1-10 keV \citep{imanishi_2003, ozawa_x-ray_2005, giardino_onset_2007}, with a typical luminosity of $\sim10^{30}$ erg.s$^{-1}$. Depending on the YSO emission spectrum and on the geometry and composition (dust and gas densities) of the protoplanetary disk, X-rays can penetrate the disk more deeply than UV photons and therefore irradiate more molecular ices \citep{agundez_chemistry_2018, walsh_molecular_2015}. As protoplanetary disks are generally formed within embedded YSO clusters, \citet{adams_background_2012} suggested that the X-ray background field originating from these clusters could also increase the X-ray flux that irradiates the molecular ices in the outer region of the disk (e.g., for $r \gtrsim$ 9-14 AU). Moreover, the stellar winds and the magnetic field structure produced by the young star can reduce the cosmic-ray flux incident on the protoplanetary disk by many orders of magnitude (e.g., at least ten orders of magnitude at 1 MeV and five orders of magnitude at 1 GeV; \citealt{cleeves_exclusion_2013}), which in that case might reduce the irradiation of interstellar ices by cosmic rays and subsequent secondary UV photons. Recently, X-ray induced photodesorption from interstellar ices has been proposed as a possible route for nonthermal desorption of H$_2$O and its photo-fragments in protoplanetary disks \citep{dupuy_x-ray_2018}. In this experimental study, an average X-ray photodesorption yield for H$_2$O from pure water ice was estimated from $\sim$10$^{-5}$ to $\sim$10$^{-3}$ molecule/photon for different regions in protoplanetary disks. X-ray induced electron-stimulated desorption (XESD) was proposed as a possible mechanism for the photodesorption of neutral molecules from water ice: following X-ray photoabsorption, the excited molecular state decays by releasing an Auger electron of $\sim$500 eV that thermalizes through the ice and creates secondary valence ionizations or excitations of the neighboring molecules, ultimately leading to their desorption at the ice surface. X-ray photodesorption in the 250-1250 eV range has also been studied for H$_2$O:CO:NH$_3$ ices \citep{jimenez-escobar_x-ray_2018}, but no methanol desorption was detected. Irradiation of a mixture of H$_2$O:CO:NH$_3$ (2:1:1) covered by a layer of CO:CH$_3$OH (3:1) with 250-1250 eV X-rays at high flux (120 minutes at $7.6 \times 10^{14}$ photon/s) did not show significant methanol desorption compared to CO, CO$_2$, HCO, and H$_2$CO desorption \citep{ciaravella_x-ray_2020}. X-ray photodesorption (at 537 eV) of ions from pure methanol ice at 55 K was also studied by \citet{andrade_x-ray_2010} : a photodesorption yield of $\sim 5 \times 10^{-9}$ ions/photon was estimated for H$_2$CO$^+$ and CH$_2$OH$^+$, for instance. Finally, X-ray photodesorption of neutral molecules from methanol-containing ices has not been constrained so far. 
\\\\
This is paper I of an experimental work dedicated to the study and quantification of the photodesorption from methanol-containing ices in the X-ray energy range. It focuses on the X-ray photodesorption from pure methanol ices with the aim to provide absolute yields for the main desorbing neutral species, including methanol itself, but also smaller fragments, or even more complex molecules, and to shed light on the involved microscopic mechanisms responsible for their ejection. The work is based on the use of the monochromatized output of the SEXTANTS beam line (SOLEIL synchrotron facility, St Aubin, France) in the O K-edge region of the CH$_3$OH molecule (525 - 570~eV). Although pure CH$_3$OH ices are not likely to be found in the ISM, studying them is the first necessary step toward studying those of more complex and astrochemically relevant binary ices containing methanol. This study is presented in paper II. 
\\\\
In paper I, section 2 introduces the experimental setup and procedures. Section 3 presents the main experimental results we obtained on the pure methanol ices, including X-ray absorption profiles and energy-resolved photodesorption yields derived from our experiments. Section 4 focuses on the molecular mechanisms at the origin of the photodesorption and on comparisons with the vacuum UV (VUV) photodesorption from pure methanol ices. Finally, section 5 presents and discusses the extrapolation of our results to the astrophysical conditions.

\section{Experimental procedures} 

\subsection{Experiments}
Experiments were conducted using the SPICES 2 setup (Surface Processes and ICES 2). It consists of an ultra-high vacuum (UHV) chamber (base pressure $\sim 10^{-10}$ mbar) within which a rotatable copper substrate (polycrystalline oxygen-free high-conductivity copper) can be cooled down to T$\sim$15 K by a closed-cycle helium cryostat. Its temperature is controlled with 0.1 K precision, and it is electrically insulated from its sample holder by a Kapton foil, allowing the measurement of the drain current generated by the electrons escaping its surface after X-ray absorption, referred to as the total electron yield (TEY) in the following. The molecular ices are formed using a dosing system: a tube positioned a few millimeters away from the substrate allows injecting a partial pressure of methanol gas on the cold substrate surface without notably modifying the base pressure in the chamber, resulting in frozen methanol ice. The ice thickness is calibrated with the temperature-programmed desorption (TPD) technique \citep{doronin_adsorption_2015}, with a relative precision of about 10 \%. The ice thickness is expressed in monolayer (ML), equivalent to a molecule surface density of $10^{15}$ molecules.cm$^{-2}$. In our experiments, we studied ices of $\sim$100 ML, deposited at 15 K.
\\\\
The X-ray photon source of the SEXTANTS beam line of the SOLEIL synchrotron facility was connected to the SPICES 2 setup to run our experiments. We used photons in the range of 525-570 eV, corresponding to the ionization edge of the O(1s) electron, with a resolution of $\Delta E = 150$ meV, where $E$ is the photon energy. The flux, measured with a calibrated silicon photodiode, was approximately $1.5 \times 10^{13}$ photon.s$^{-1}$, with little variation except for a significant dip around 534 eV due to the O 1s absorption of oxygen pollution on the optics of the beam line. The beam was sent at a $47^{\circ}$ incidence on the ice surface in a spot of $\sim$0.1 cm$^2$.
\\\\
While the ices were irradiated, the photodesorption of neutral species was monitored by recording the desorbed molecules in the gas phase using a quadripolar mass spectrometer (QMS). Each gas-phase species was probed by monitoring the mass signal of its corresponding intact cation: for example, CH$_3$OH and HCOOH were recorded by selecting mass 32 and 46, respectively. Different irradiation procedures were made :
\\\\
- short irradiations at 534, 541, and 564 eV were conducted to measure the photodesorption yields at these fixed energies. The irradiation time per fixed energy was approximately 10 seconds (higher than the acquisition time of the QMS, which is 100 ms). This mode allowed us to probe the photodesorption from the ices with a relatively low irradiation fluence, mainly to prevent the photo-aging of the condensed systems from having a significant effect on the detected signals: the fluence received by the ice during these short irradiations is $\sim 5 \times 10^{15}$ photon/cm$^{2}$.
\\\\
- continuous irradiations from 525 to 570 eV in steps of 0.5 eV (scans) allowed us to record the photodesorption spectra as a function of the photon energy. The irradiation time per scan was approximately 10 minutes and the fluence received by the ice was $\sim 1 \times 10^{17}$ photon/cm$^{2}$. During these irradiations, the TEYs were also measured as a function of the photon energy with a scan step of 0.5 eV.
\\\\
These irradiations were performed successively on the same sample, allowing us to explore the aging effect on the extracted photodesorption yields. Finally, at the end of these irradiation experiments, TPD experiments (from 15 K to 200 K) were used to evaporate all the molecules from the substrate surface before a new ice was formed.

\subsection{Calibration of the photodesorption yields per incident photon}
In order to compute the photodesorption yields (in molecule desorbed per incident photon, displayed as molecule/photon for more simplicity in the following) from the QMS signals, the following procedure was conducted:
\\\\
- the signal given by the QMS was corrected for the photon flux profile and for the apparatus function of the QMS for a given mass to take the transmission and detection efficiency of the QMS into account.
\\\\
- using TPD experiments \citep{doronin_adsorption_2015}, we determined a proportionality factor $k_X$ between the molecular desorption flux and the QMS signal. This calibration was made on CH$_3$OH where the difference between the monolayer and multilayer thermal desorption regime is quite clear and thus allows depositing a single monolayer of methanol and determining the corresponding integral mass signal. Using this technique, we assumed that the proportionality factor for photodesorbed methanol is similar to the factor for thermal desorption, assuming a similar angular distribution of the desorbates. This method has previously been employed in the case of X-ray photodesorption from water ices \citep{dupuy_x-ray_2018} and was confirmed in the UV range, where an infrared calibration procedure can also be used for systems with little photodissociation, such as pure CO ice.
\\\\
- the photodesorption yields for the other neutral species, such as fragments for which TPD calibration is not possible, can be deduced from the CH$_3$OH calibration by taking into account the differences in electron-impact ionization cross sections and apparatus functions of the mass filter. This results in using the following formula \citep{dupuy_spectrally-resolved_2017}: 
\begin{align}
k_X = \dfrac{\sigma(X^+/X)}{\sigma(CH_3OH^+/CH_3OH)} \times \dfrac{AF(X)}{AF(CH_3OH)} \times k_{CH_3OH}
\end{align}
where $X$ is the neutral species considered, $AF(X)$ is the apparatus function of our QMS for the given species, $k_X$ is its proportionality factor, and $\sigma(X^+/X)$ is the electron-impact ionization cross section for the $X$ neutral species, taken at 70 eV. The values for these different cross sections were found in the literature \citep{Freund_1990,Srivastava_1996, Straub_1998, Joshipura_2001, Liu_2006,vacher_electron_2009, zawadzki_electron-impact_2018}.
\\\\
- for molecules that can originate from the cracking of their parent molecules in the QMS, we used the cracking patterns available on the National Institute of Standards and Technology (NIST) chemistry Webbook to correct the signals. More importantly, the cracking of CH$_3$OH into CH$_2$OH$^+$ and/or CH$_3$O$^+$ was used to compute the photodesorption yield of methanol (see section 3).
\\\\
- uncertainties on the photodesorption yields were computed by taking the signal-to-noise ratio for the corresponding mass channel of our QMS into account.

\subsection{Photodesorption yields per absorbed photon}

In section 4 we compute the X-ray photodesorption yields per absorbed photon for pure methanol ice, which is an interesting figure with which to discuss the photodesorption mechanisms using the following formula:
\begin{ceqn}
\begin{align}
\Gamma^{abs} = \dfrac{\Gamma^{inc}}{1-e^{-\sigma N}}
\end{align}
\end{ceqn}
where $\sigma$ is the photoabsorption cross section, $N$ is the column density of molecules that we consider to be involved in the photodesorption process, and $\Gamma^{inc}$ is the photodesorption yield per incident photon derived from our experiments at a given energy. We assumed that the X-ray absorption cross section of condensed phase methanol is equal to that in gas phase at 564 eV, which is $\sim$0.52 Mbarn \citep{ISHII198855}. This assumption appears to be reasonable as the X-ray photoabsorption of methanol near the O K edge is due to core electrons whose electronic structure should not be significantly modified when it changes from gas phase to condensed phase. To our knowledge, no experiments concerning the study of the column density involved in the X-ray photodesorption process from methanol-containing ice have been carried out. Thus we assumed that the ice thickness involved in the photodesorption from methanol-containing ice is governed by the radius of the electron cloud created by the Auger electron and is similar to the involved thickness for water ice, which is 10 nm \citep{timneanu_auger_2004} and corresponds to approximately 30 ML. Finally, we considered a molecule surface density of $10^{15}$ molecule.cm$^{-2}$. 

\subsection{Photodesorption yields for astrophysical models}
Our experimental photodesorption spectra are limited to the 525-570 eV range, while X-ray emission spectra of YSOs are typically in the 0.1-10 keV range. In order to provide photodesorption yields for astrochemical models, we extrapolate our results in section 4 to higher photon energies. We observed that the X-ray photodesorption from our ices is following the absorption of the O 1s core electrons in the 525-570 eV range (see section 3.1). Above 570 eV, the ionization threshold of O 1s core electrons is exceeded and the X-ray absorption is dominated by O 1s ionization, also resulting in the release of an Auger electron followed by a cascade of low-energy secondary electrons within the ice. We can therefore extrapolate our experimental photodesorption spectra above 570 eV by assuming that they follow the same variations as the X-ray O 1s ionization cross section, which is assumed to be similar to that of gas-phase methanol \citep{Berkowitz:1087021}. For the 525-570 eV range, the photodesorption spectra follow the TEY, starting from the estimated yields at 564 eV, as observed in section 4. Then, for a given X-ray emission spectrum $\phi(E)$, where $E$ is the photon energy, we can estimate an average photodesorption yield $Y_{avg}$ using the following formula:
\begin{ceqn}
\begin{align}
Y_{avg} = \dfrac{\int \Gamma^{inc}(E)  \phi(E)  dE}{\int \phi(E)  dE}
\end{align}
\end{ceqn}
where $\Gamma^{inc}(E)$ is our measured photodesorption yield. In protoplanetary disks, as X-rays are attenuated by dust and gas depending on the region considered, we also provide an average photodesorption yield for attenuated X-ray spectra $\phi^{att}(E)$ corresponding to different H column densities using the Beer-Lambert law,
\begin{ceqn}
\begin{align}
\phi^{att}(E) = \phi(E) e^{-\sigma_{pe}(E) n_H}
\end{align}
\end{ceqn}
where $n_H$ is the H column density and $\sigma_{pe}(E)$ is the photoelectric cross section of gas and dust in a typical T Tauri protoplanetary disk estimated by \citet{bethell_photoelectric_2011}.

\begin{figure} [h!]
\resizebox{\hsize}{!}{\includegraphics{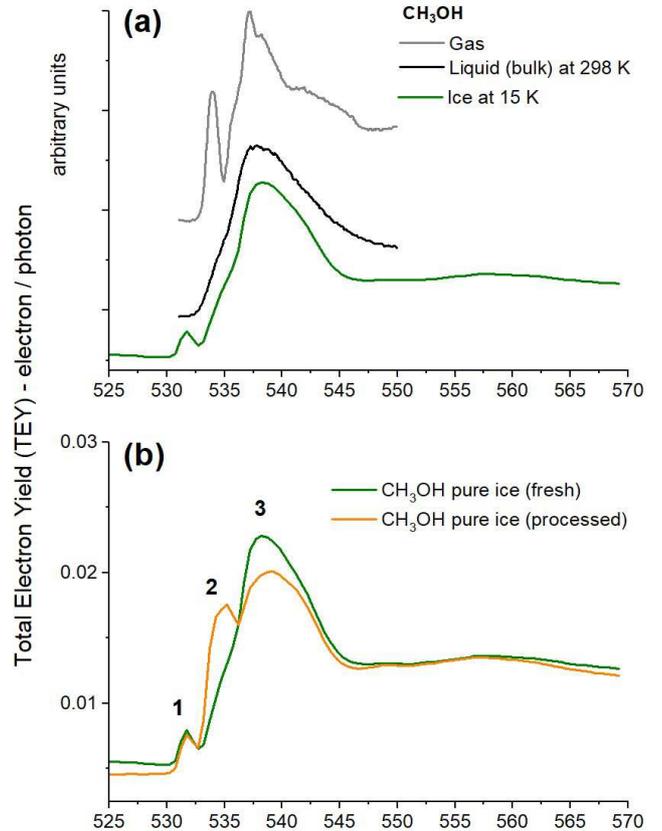}}
\caption{(a) TEY as a function of photon energy for methanol in gas phase, in liquid microjets at 298 K \citep{wilson_x-ray_2005}, and in condensed phase at 15 K (measured in our experiments). (b) TEY of pure methanol ice, as deposited (fresh ice), and after having been processed by an irradiation fluence of 1.10$^{17}$ ph/cm$^{2}$ (processed ice).}
\label{TEY}
\end{figure}

\begin{figure*}
\centering
\includegraphics[width=17cm]{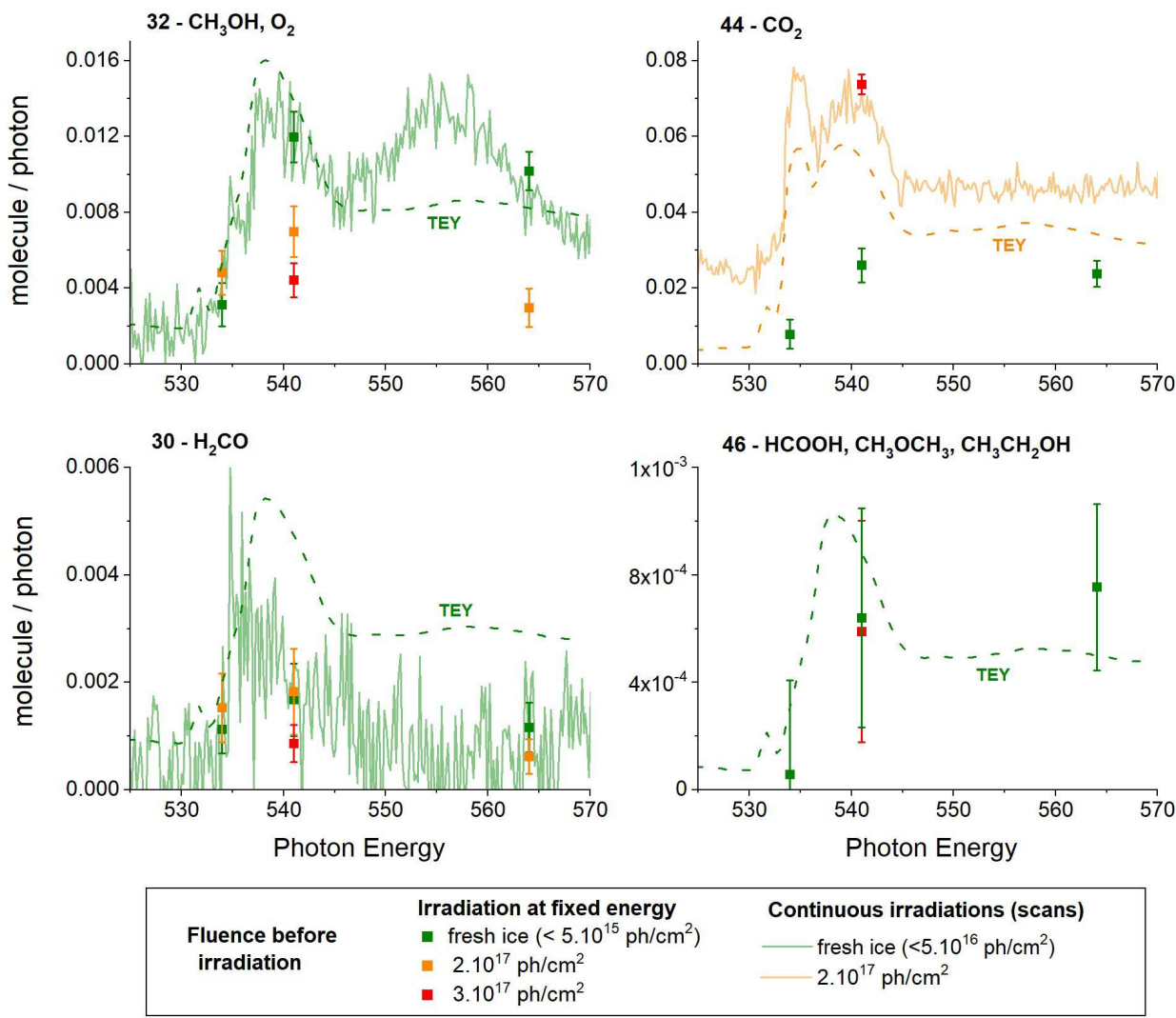}
\caption{Photodesorption spectra for masses 30, 32, 44, and 46 in molecule/photon from pure methanol ice at 15 K, with the associated molecules. The measurements at fixed energy are represented by the squares with error bars. The scan experiments are represented by solid lines. The TEYs measured during the scan experiments are also shown as dashed lines in arbitrary units. Information about the fluence received by the ice before each measurement is displayed in the bottom panel.}
\label{Graph_Pur}
\end{figure*}

\section{Results} 
\subsection{X-ray absorption of the ices and TEY} 
The TEY measured from our experiments are shown in Figure \ref{TEY}.(b). The different features observed are labeled and can be assimilated to the X-ray absorption of the ice: after a photon is absorbed by a core O 1s electron, the decay of the resulting molecular excited state leads to the release of an Auger electron of $\sim$500 eV. The thermalization of this Auger electron by inelastic scattering within the ice creates secondary valence excitations and ionizations of neighboring molecules, leading to a cascade of secondary electrons. The current generated by the escape of these electrons from the ice surface can be quantified per incident photon and provides information about the core electronic structure of O-bearing molecules in condensed phase near the O K edge.
\\\\
In Figure \ref{TEY}.(a) we compare our TEY measurement for pure methanol ice (100 ML, at 15 K) with X-ray absorption spectroscopy of gas-phase methanol and liquid methanol microjets \citep{wilson_x-ray_2005}. The first two peaks of the gas-phase methanol (at 534 eV and 537 eV) are attributed to 3s and 3p Rydberg orbitals, with some $\sigma^*$ character of the O-H and C-O bonds, respectively. From the gas phase to the liquid phase, these peaks are broadened as a result of the greater orbital overlap with neighboring molecules in the liquid state, and the gas-phase first peak is also blueshifted by $\sim$1 eV. No significant differences are observed between the TEY of liquid methanol at 298 K and methanol in condensed phase at 15 K, except for the small peak at 531.5 eV that is labeled peak 1 in Figure \ref{TEY}.(b). We lack published data on the X-ray absorption of solid methanol in the considered energy range, therefore it is difficult to propose an attribution for this feature. Its low dependence on the irradiation fluence indicates a structure that is associated either with intact methanol ice, that is, not to a photoproduct, or with the substrate itself. However, the mean depth reachable by the photoelectrons is estimated to be about 30 ML (see section 2.3). This is three times lower than the methanol-ice thickness, which makes a possible contribution of the substrate to the TEY negligible. Finally, the broad peak near 537 eV (labeled 3 in Figure \ref{TEY}.(b)) decreases when the ice is irradiated, showing that the destruction of CH$_3$OH occurs in the ice. From the photodesorption yields that are presented in the next sections, we can estimate that about 1 ML of methanol is photodesorbed for a fluence of $10^{17}$ photon/cm$^{2}$. The ice thickness is 100 ML and the probed thickness is $\sim 30$ ML, therefore the TEY is not expected to be strongly affected by the photodesorption. Instead, we attribute the decrease in peak 3 to the photodissociation of the methanol and to photo-chemistry. This is further confirmed by the appearance of peak 2 with an ongoing irradiation fluence ($> 1 \times 10^{17}$ photon/cm$^{2}$). This peak can be attributed to the 1s$^{-1} \pi^*$ resonance (1$\sigma$-2$\pi$ transition) of CO molecules in condensed phase, which are formed by X-ray induced chemistry. This feature is also observed for CO in gas phase \citep{puttner_vibrationally_1999} and for pure CO ice \citep{dupuy:tel-02354689}. Similar photoaging effects have been observed in X-ray irradiation experiments of pure methanol ice \citep{laffon_photochemistry_2010, chen_soft_2013}.

\subsection{Photodesorption yields}

\begin{table*}[h!]
\caption{X-ray photodesorption yields (in $10^{-3}$ molecule/photon) at 564 eV from pure methanol ice at 15 K and for a dose between 5.10$^{15}$ and 2.10$^{16}$ photon/cm$^{2}$.}
\begin{center}
\begin{threeparttable}[h!]

\begin{tabular}{p{4,5cm}p{2,1cm}p{2cm}p{5cm}p{2,1cm}}
\hline \hline \\
Mass - photodesorbed species &  \centering Yield & & Mass - photodesorbed species & Yield 
\\[0,2cm]\hline\\

15 - CH$_3$\tnote{(1)} &  \centering $1.3^{\pm0.2}$ & & 30 - H$_2$CO & $1.2^{\pm0.5}$ \\ [0,2cm]

16 - CH$_4$, O & \centering $1.1^{\pm0.2}$ & & 32 - CH$_3$OH & $7.6^{\pm0.9}$ \\ [0,2cm]

17 - OH & \centering ND\tnote{*} & & 44 - CO$_2$ & $24^{\pm1}$ \\ [0,2cm]

18 - H$_2$O & \centering $7.5^{\pm1.1}$ &  & 46 - HCOOH, C$_2$H$_6$O isomers & $0.8^{\pm0.3}$ \\ [0,2cm]

28 - CO & $(1.2)^{\pm0.1} \times 10^{2}$ \\ [0,2cm]
\hline

\end{tabular}

\begin{tablenotes}
            \item[*] ND = Not detected: the desorption signal measured for the considered species is lower than our signal-to-noise ratio, meaning that we did not detect its desorption from the ice. Considering the noise profile in the mass channel 17 of our QMS, if OH photodesorption occurs, the photodesorption yield is $< 5 \times 10^{-4}$ molecule/photon.               
                \item[(1)] The yield is given without being able to correct for the cracking of CH$_4$ into CH$_3$ in the QMS. These values may therefore be overestimated. 
\end{tablenotes}
\end{threeparttable}
\end{center}
\label{Yields}
\end{table*}

In Figure \ref{Graph_Pur} we report the photodesorption yields from pure methanol ice in molecule desorbed per incident photon (displayed molecule/photon for more simplicity) we derived from our measurements. We do not display all available data for more clarity, as we did not observe behaviors different from the data we present. The TEY measurements are also shown in arbitrary units (only the energy dependence is of interest when comparing with the photodesorption). The photodesorption yields derived from the irradiations at fixed energy are consistent with those measured during the scan experiments, except when the fluence received by the ice differs. The reason is that the induced chemistry occurs in the ice. We discuss this in the next sections. The remaining relevant data we obtained are summarized in Table \ref{Yields}, where the yields are derived from our fixed energy experiments. As a lower fluence is used compared to the scan experiments, the aging effect is limited for these yields: the fluence received by the ice before measurement ranges from $5 \times 10^{15}$ to $2 \times 10^{16}$ photon/cm$^{2}$. These yields are derived at a fixed energy of 564 eV.
\\\\
CO and CO$_2$ are the most strongly desorbing species with a photodesorption yield at 564 eV of $\sim$ 0.1 molecule/photon and $\sim$ 0.02 molecule/photon, respectively. OH X-ray photodesorption is not detected during irradiations. For the desorption signal on the mass 16, we were not able to distinguish between $^{12}$CH$_4$ or atomic O. For the mass 15, we give the raw data that are not corrected for any cracking pattern (especially from the mass 16, which could be attributed to CH$_4$ photodesorption) and it may be overestimated.
\\\\
As the QMS signal registered on the mass 32 could correspond to O$_2$ or CH$_3$OH desorption, we used the signal on the mass 31 to estimate the weight of CH$_3$OH photodesorption on the mass channel 32. We assumed that the desorption of CH$_2$OH or CH$_3$O radical (that would contribute to the mass channel 31) is negligible in our experiments with pure methanol ice, so that the signal on the mass 31 only originates from the cracking of desorbing CH$_3$OH into CH$_2$OH or CH$_3$O (which are the main fragments) in the ionization chamber of the QMS. This hypothesis agrees well with our conclusions about the photodesorption mechanisms at play in section 4.1 and appears reasonable considering the data available from similar experiments: the non-negligible formation of CH$_2$OH is clearly visible by infrared spectroscopy \citep{chen_soft_2013} when pure methanol ice is irradiated at 550 eV at similar fluence, but UV photodesorption of CH$_2$OH or CH$_3$O was not observed ($< 3 \times 10^{-6}$ molecule/photon; \citealt{bertin_uv_2016}) from pure methanol ice. We then found that a maximum of $\sim$75\% of the signal on the mass 32 could correspond to CH$_3$OH photodesorption (after correction for the cracking of CH$_3$OH into CH$_2$OH or CH$_3$O, this brings the signal on the mass 31 to below our detection threshold, which is $5 \times 10^{-4}$ molecule/photon), at 541 and 564 eV and for fresh ice (fluence $< 5 \times 10^{15}$ photon/cm$^2$). The remaining signal on the mass 32 ($\sim$25\%) then originates from O$_2$ photodesorption, whose yield at 564 eV is then $\sim 2.5 \times 10^{-3}$ molecule/photon. When the fluence increased to 3.10$^{17}$ photon/cm$^2$ (see Figure \ref{Graph_Pur}), we found that only $\sim$30\% of the signal on the mass 32 could correspond to CH$_3$OH photodesorption at 541 eV.
\\\\
We also observed a photodesorption signal on the mass channel 46, which can be attributed to either HCOOH (formic acid) and/or C$_2$H$_6$O isomers (ethanol and dimethyl ether). These molecules have been detected by infrared spectroscopy when pure methanol ice at 14 K was irradiated with X-rays of 550 eV \citep{chen_soft_2013}. 

\section{Discussion}
\subsection{Mechanisms for the X-ray photodesorption of neutrals from pure methanol ice}
Figure \ref{Graph_Pur} shows a correlation between the photodesorption spectra from pure methanol ice and the TEY as a function of the photon energy (except for the broad peak observed on the photodesorption spectrum of the mass channel 32 between 545 and 565 eV, which is due to an unstable background noise and does not reflect any particular physical mechanism). The photodesorption process is thus linked to the X-ray absorption of the ice and could be caused by two main mechanisms. The first mechanism is direct desorption, which means that the desorption occurs after the decay of an excited molecular state due to the photoabsorption on the ice surface. The second mechanism is called X-ray induced electron-stimulated desorption (XESD). In this case, the desorption originates from the multiple events caused by the cascade of secondary electrons generated by the X-ray absorption of O 1s core electrons within the ice followed by the thermalization of the Auger electron. It is not clear whether direct desorption or XESD dominates the desorption process because the fact that the photodesorption spectra follow the TEY is compatible with either process. However, in the following discussion, we develop arguments in favor of XESD. For simpler molecular ices such as water ice or CO ice, X-ray photodesorption experiments \citep{dupuy_x-ray_2018, dupuy:tel-02354689} have demonstrated that XESD was the dominant process.
\\\\
When the fluence received by our pure methanol ice is increased, the TEY and photodesorption spectrum shapes are modified in a similar way. The main modification is that a peak near 535 eV arises (see Figure \ref{TEY} or \ref{Graph_Pur}) that is as high as the broad peak near 540 eV that we attributed to CH$_3$OH molecules. It can be explained by the formation of CO inside the ice as a result of X-ray induced chemistry. This means that the photodesorption process can originate from either CO (534 eV) or CH$_3$OH (541 eV) X-ray photoabsorption and lead to quite similar photodesorption efficiency at these energies for a processed ice. In Figure \ref{Graph_Pur}, for instance, the photodesorption yields on the mass channel 32 at 534 eV and at 541 eV are $0.5 \times 10^{-2}$ molecule/photon and $0.7 \times 10^{-2}$ molecule/photon, respectively, for a fluence of $2 \times 10^{17}$ photon/cm$^2$ , whereas they differ by a factor of $\sim 4$ for fresh ice. This means that when sufficient CO molecules are present in the ice, CO X-ray absorption at 534 eV leads to the same photodesorption efficiency of the mass 32 as CH$_3$OH X-ray absorption at 541 eV. This is also observed for CO$_2$ photodesorption yields at 534 eV and 541 eV, which are $7.6 \times 10^{-2}$ molecule/photon and $6.8 \times 10^{-2}$ molecule/photon, respectively, for a fluence of $2 \times 10^{17}$ photon/cm$^2$ , whereas they differ by a factor of $\sim 4$ for fresh ice. This highlights the importance of the ice composition and shows that the X-ray photodesorption process could be an indirect process where the photodesorption of one molecule is triggered by the X-ray absorption of another one, which is an argument in favor of an XESD process.
\\\\
When we now compare the photodesorption yields at 541 eV between fluences $< 5 \times 10^{15}$ photon/cm$^2$ and at $3 \times 10^{17}$ photon/cm$^2$ in Figure \ref{Graph_Pur}, the CO$_2$ photodesorption first increases from $2.6 \times 10^{-2}$ molecule/photon to $7.3 \times 10^{-2}$ molecule/photon. We also observed this phenomenon for CO photodesorption yield (the data are not shown for more clarity), which increased from $1.9 \times 10^{-2}$ molecule/photon to $3.5 \times 10^{-2}$ molecule/photon. Second, the estimated yield for the X-ray photodesorption of CH$_3$OH from pure methanol ice decreased by almost one order of magnitude from $9.0 \times 10^{-3}$ molecule/photon to $1.3 \times 10^{-3}$ molecule/photon. This indicates that the photodesorption of CH$_3$OH is higher for a lower fluence received by the ice when more intact methanol molecules are present in the ice. This aging process favors the photodesorption of simpler molecules such as CO$_2$ or CO. \citet{laffon_photochemistry_2010} estimated with NEXAFS spectroscopy (at the C K-edge) that X-ray irradiation at 150 eV of pure methanol ice at 20 K leads to a survival rate of 50\% for methanol after an absorbed dose of 1.1 MGy. In our fixed-energy experiments, we irradiated pure methanol ice with fluences between $5 \times 10^{15}$ photon/cm$^{2}$ and $2 \times 10^{16}$ photon/cm$^{2}$. Because we irradiated a volume of 0.1 cm$^{2} \times$100 ML, with a mean energy of $\sim$550 eV, and when we consider a volumic mass of condensed methanol of $\sim 0.64$ g.cm$^{-3}$ (at 20 K; \citealt{luna_densities_2018}) and an X-ray absorption cross section of $\sim$0.6 Mbarn \citep{ISHII198855}, the absorbed doses used in our fixed energy experiments change from $\sim 2$ MGy to $\sim 15$ MGy, which is quite similar to the absorbed doses in \citet{laffon_photochemistry_2010}. This indicates that we could expect a methanol destruction rate of about 50\% for our low-fluence experiments. In similar experiments, when irradiating a H$_2$O:CH$_4$:NH$_3$ (2:1:1) ice mixture covered by a layer of CO:CH$_3$OH (3:1) with 250-1250 eV X-rays during 120 minutes with a flux of $7.6 \times 10^{14}$ photon/s, higher by almost two order of magnitudes than our experiments, \citet{ciaravella_x-ray_2020} did not detect a desorption signal on the mass channel 31 (attributed to methanol desorption) and estimated that only $\sim$ 20\% of methanol molecules remained intact in the first minutes of the irradiation. The irradiation flux therefore appears to be critical for detecting methanol desorption in X-ray irradiation experiments of methanol-containing ices. A lower X-ray flux appears to favor methanol desorption because the methanol destruction rate is lower. This destruction of methanol molecules could also have a significant effect on the formation and desorption of more complex molecules. 
\\\\
In Table \ref{UV_X} we compare the photodesorption yields per absorbed photons in the X-ray (at 564 eV, from Table \ref{Yields}) and the UV domains (at 10.5 eV; from \citet{bertin_uv_2016}) for CO, H$_2$CO, and CH$_3$OH from pure methanol ice using the method described in section 2.3. To compute the UV photodesorption yield per absorbed photon, we used the same formula as for the X-ray photodesorption, and we considered the UV photodesorption yields measured at 10.5 eV in \citet{bertin_uv_2016}, which are $1.5 \times 10^{-4}$, $1.25 \times 10^{-5}$ and $1.5 \times 10^{-5}$ molecule per incident photon for CO, H$_2$CO, and CH$_3$OH, respectively. The column density involved in the UV photodesorption is well constrained for pure CO ices \citep{bertin_uv_2012} and equal to $\sim$3 ML. We took the same value for pure methanol ice. Finally, the UV photoabsorption cross section of methanol in condensed phase is directly available in the literature, and is equal to 8.6 Mbarn at 10 eV \citep{cruz-diaz_vacuum-uv_2014}. \\

\begin{table}[!h]
\caption{Photodesorption yields in molecule per absorbed photon in UV (at 10.5 eV) and X-ray domains (at 564 eV) from pure methanol ice.}
\begin{threeparttable}
\centering
\begin{tabular}{p{2.2cm}|p{1.5cm}p{1.7cm}p{1.7cm}}
\hline \hline \\
Ice at 15 K & Molecule & X-rays & UV\tnote{(1)}  
\\[0,2cm]\hline\\

\multirow{3}{*}{Pure CH$_3$OH}  &  CO  &  8.0 &  $5.9 \times 10^{-3}$ \\[0,1cm] 

&   H$_2$CO  &  0.08 &  $4.9 \times 10^{-4}$ \\[0,1cm] 
 
&   CH$_3$OH  &  0.49 &  $5.9 \times 10^{-4}$ \\[0,2cm]
  
\hline    
\end{tabular}
\begin{tablenotes}
            \item[(1)] from \citet{bertin_uv_2016}
\end{tablenotes}         
\end{threeparttable}
\label{UV_X}
\end{table}

The photodesorption yields per absorbed photons in Table \ref{UV_X} show that the X-ray yields are approximately three orders of magnitude higher than the UV yields. This difference is still significant in the photodesorption yields per absorbed eV: as X-rays carry $\sim$500 eV and UV photons carry $\sim$10 eV, the CH$_3$OH X-ray and UV photodesorption yield is $1.0 \times 10^{-3}$ molecule/absorbed eV and $5.9 \times 10^{-5}$ molecule/absorbed eV, respectively, which gives a X-ray yield higher than a UV yield by more than one order of magnitude. This is also the same difference as for CO X-ray and UV photodesorption yields, which are $1.6 \times 10^{-2}$ molecule/absorbed eV and $5.9 \times 10^{-4}$ molecule/absorbed eV, respectively. For H$_2$CO, this difference is smaller than one order of magnitude: H$_2$CO X-ray and UV photodesorption yields are $1.6 \times 10^{-4}$ molecule/absorbed eV and $4.9 \times 10^{-5}$ molecule/absorbed eV, respectively. 
\\\\
For pure methanol ice, X-ray photodesorption per absorbed eV is therefore more efficient by one order of magnitude than UV photodesorption. This could be explained by a difference in the induced chemistry. The processing of pure methanol ice by high-energy electrons (of $\sim$ 1 keV) does not appear to depend on the incident electron energy \citep{bennett_mechanistical_2007, mason_electron_2014}, indicating that the chemistry is driven by low-energy secondary electrons resulting from the thermalization of the incident high-energy electron within the ice. A similar process is expected for X-ray induced chemistry, where low-energy secondary electrons are created by the thermalization of the Auger electron. A comparison of the chemistry induced by high-energy electrons, X-rays, and UV photons could provide some indications about the difference observed in the photodesorption efficiency and the mechanisms at play. The processing of pure methanol ice by UV photons (ice at 20 K irradiated by 7–10.5 eV photons; \citealt{oberg_photochemistry_2016}), X-rays (ice at 14 K irradiated by monochromatic X-rays at 300 eV and 550 eV and by broadband 250–1200 eV X-rays; \citealt{chen_soft_2013}), and high-energy electrons (ice at 11 K irradiated by 5 keV electrons; \citealt{bennett_mechanistical_2007}) leads to similar photo-products within the ice. However, there are some slight differences that may be important. For example, kinetic modeling of induced chemistry shows some differences in the dissociation branching ratios of CH$_3$OH between UV \citep{oberg_photochemistry_2016} and electron irradiations \citep{bennett_mechanistical_2007}. Moreover, \citet{chen_soft_2013} compared the number of molecules produced per absorbed eV in pure methanol ice between UV (from \citet{oberg_photochemistry_2016}), X-rays, and 5 keV electron (from \citet{bennett_mechanistical_2007}) experiments and suggested a lower product efficiency by UV irradiation than by X-rays and 5 keV electron irradiations. These differences between UV and secondary low-energy electron induced chemistry may translate into a difference between X-ray and UV photodesorption efficiency for pure methanol ice that favors the X-ray efficiency over the UV efficiency (Table \ref{UV_X}). 
\\\\
The way in which the energy is deposited within the ice and its consequence on the induced chemistry could provide an interesting route for explaining X-ray and UV photodesorption efficiency from pure methanol ice. When an X-ray is absorbed in the ice, most of the energy ($\sim$ 500 eV) goes into the Auger electron, and the deposited energy is spatially localized around the thermalization path of this Auger electron. Depositing an equivalent amount of energy ($\sim$ 500 eV) with UV photons in the ice may yield a different spatial distribution of the total deposited energy as it is expected to be homogeneously distributed in the ice. At the low temperatures considered here (T $\sim$ 15 K), where the diffusion of molecules is limited, this localization of the deposited energy would therefore favor reactions between neighboring excited molecules, radicals or ions in the case of X-ray absorption compared to UV absorption. Moreover, X-ray induced chemistry could involve a richer reaction network as a result of ion chemistry because molecular ionizations are easily produced in the thermalization path of the Auger electron, whereas UV photons in the range of 7-14 eV are expected to produce fewer ions because their energy is close to the ionization energy of CH$_3$OH.
\\\\
Finally, for the specific case of CH$_3$OH photodesorption, \citet{bertin_uv_2016} suggested that the UV photodesorption of methanol from pure methanol ice originates from the exothermic recombination of CH$_3$O/CH$_2$OH into CH$_3$OH followed by its desorption (also suggested by \citet{oberg_photochemistry_2016}). An exothermic recombination was also proposed in similar experiments as a possible route for the UV photodesorption of H$_2$CO from formaldehyde-containing ices \citep{feraud_vacuum_2019}. \citet{bennett_mechanistical_2007} suggested that the main dissociation channels of CH$_3$OH in condensed phase by 5 keV electrons lead to CH$_2$OH, CH$_3$O, and CH$_4$ formation. CH$_2$OH/CH$_3$O reactions with H atoms were also suggested as a recombination channel to re-form methanol. This recombination process could be responsible for methanol X-ray photodesorption from pure methanol ice. Moreover, the reaction CH$_3$OH $\xrightarrow{4e^-}$ CO + 4H in condensed methanol is very efficient for low-energy electrons \citep{lepage_low_1997}, as suggested in \citet{laffon_photochemistry_2010}, which is certainly why a significant CO formation is observed in the TEY evolution with increasing fluence in our experiments. Thus, as X-ray induced chemistry may produce a large amount of H atoms in pure methanol ice, the recombination of CH$_2$OH and/or CH$_3$O with H atoms to re-form methanol, followed by its desorption, should be more efficient than the same process for UV-irradiated pure methanol ice, and this recombination process should be favored by a spatially localized X-ray induced chemistry, as explained before. This supports our finding that the CH$_3$OH photodesorption yield/absorbed eV from pure methanol ice is larger for X-rays than for UV photons and is consistent with our previous results that stated a lower X-ray photodesorption efficiency of methanol when fewer intact methanol molecules are present in pure methanol ice as a result of an increasing X-ray fluence.

\subsection{Astrophysical implications}
\begin{figure} [h!]
\centering
\includegraphics[width=7cm]{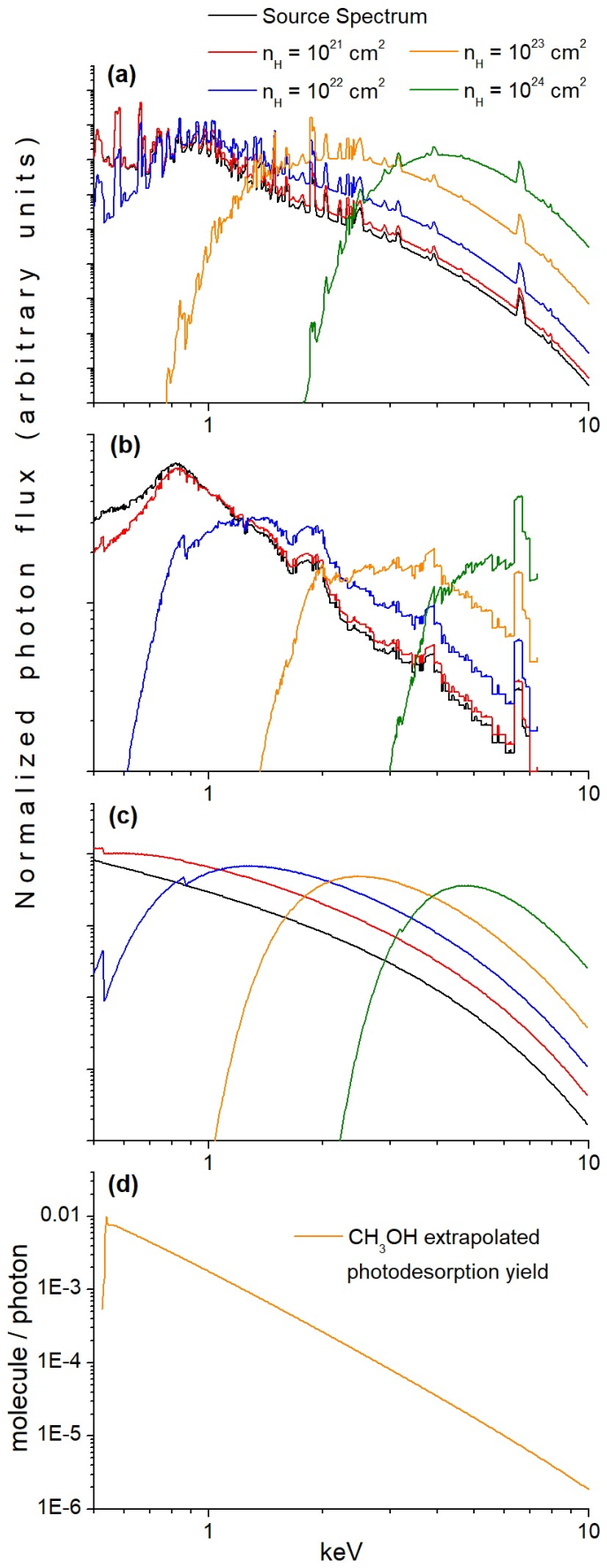}
\caption{Normalized (with respect to the area) X-ray spectra of (a) TW Hya \citep{nomura_molecular_2007}, (b) Herbig Ae HD 104237 \citep{skinner_new_2004}, and (c) X-ray background field (XBGF) of YSO clusters \citep{rab_x-ray_2018}. The source spectrum is represented along with its attenuation for different H column densities. (d) Extrapolated photodesorption yield spectra from our experimental results and using the absorption cross section of gas-phase methanol from \citet{Berkowitz:1087021}}
\label{Astro_Graph}
\end{figure}

\begin{table*}[!t]
\caption{Average astrophysical photodesorption yield of methanol from pure methanol ice at 15 K in molecule/photon for different YSO X-ray spectra and for different attenuation n$_H$ column density, computed as detailed in section 2. The X-ray spectra used are presented in Figure \ref{Astro_Graph}.}

\begin{center}
\begin{threeparttable}

\begin{tabular}{p{3cm}p{3cm}p{3cm}p{3cm}}

\hline \hline \\
 & TW Hya\tnote{(1)} & Herbig Ae\tnote{(2)}  &  XBGF \tnote{(3)} 
\\[0,2cm]\hline\\

Source spectrum  &  3.9 $\pm$ 2.0 $\times 10^{-3}$  &  1.7 $\pm$ 0.8 $\times 10^{-3}$ &  2.4 $\pm$ 1.2 $\times 10^{-3}$ \\[0,1cm] 

n$_H$ = 10$^{21}$ cm$^2$ & 3.6 $\pm$ 2.3 $\times 10^{-3}$  &  1.5 $\pm$ 0.7 $\times 10^{-3}$ &  2.1 $\pm$ 1.0 $\times 10^{-3}$ \\[0,1cm]
 
n$_H$ = 10$^{22}$ cm$^2$  &  1.8 $\pm$ 0.9 $\times 10^{-3}$  &  5.7 $\pm$ 2.8 $\times 10^{-4}$ &  7.1 $\pm$ 3.5 $\times 10^{-4}$ \\[0,1cm]
  
n$_H$ = 10$^{23}$ cm$^2$  &  2.4 $\pm$ 1.2 $\times 10^{-4}$  &  7.9 $\pm$ 4.0 $\times 10^{-5}$ &  1.1 $\pm$ 0.5 $\times 10^{-4}$ \\[0,1cm]

n$_H$ = 10$^{24}$ cm$^2$  &  3.2 $\pm$ 1.6 $\times 10^{-5}$  &  1.5 $\pm$ 0.7 $\times 10^{-5}$ &  1.9 $\pm$ 1.0 $\times 10^{-5}$ \\[0,2cm]

\hline
                  
\end{tabular}
\begin{tablenotes}
            \item[(1)]X-ray spectrum from \citet{nomura_molecular_2007}
            \item[(2)] X-ray spectrum of Herbig Ae star HD 104237 from \citet{skinner_new_2004}
            \item[(3)] X-ray spectrum from \citet{rab_x-ray_2018}
\end{tablenotes}

\end{threeparttable}
\end{center}   
\label{Astro_Yields}

\end{table*}

In the previous section, we have derived the photodesorption yields at fixed energy under our experimental conditions. In this section, we extrapolate these yields to astrophysical environments by applying the method described in section 2, considering three different X-ray emission spectra: one of a Classical T Tauri star (CTTS), TW Hya, from \citet{nomura_molecular_2007}, one of a Herbig Ae star, HD 104237, from \citet{skinner_new_2004}, and one that represents the X-ray emission spectrum of a cluster of YSOs, computed by \citet{rab_x-ray_2018}. As these X-rays are attenuated by the disk material (gas and dust) when they reach the cold regions of the disk, we multiplied them by an attenuation factor (from \citet{bethell_photoelectric_2011}) depending on the H column density. The resulting spectra shown in Figure \ref{Astro_Graph}.(a), (b), and (c) thus represent an estimate of the local X-ray field for different regions of the disk. In Figure \ref{Astro_Graph}.(d) we reproduce the photodesorption spectrum of CH$_3$OH from 0.525 keV to 0.570 keV by starting from the estimated yield from pure methanol ice in Table \ref{Yields} at 564 eV ($7.6 \times 10^{-3}$ molecule/photon) and by taking the same variations as the TEY in the 525-570 eV range, according to what we discussed in previous sections. From 0.570 keV to 10 keV, we assumed that the photodesorption spectrum followed the X-ray absorption cross section of condensed phase methanol, the latter assumed to be similar to the X-ray absorption cross section of gas-phase methanol linked to the O 1s ionization (taken from \citet{Berkowitz:1087021}) as it is related to a core absorption and should not be significantly modified between gas phase and condensed phase. The final computations are presented in Table \ref{Astro_Yields}, where we display the estimated astrophysical average photodesorption yields of methanol in different regions of protoplanetary disks. As these yields are computed in photodesorbed molecule per incident photon, the differences in X-ray luminosity between YSOs could play an important role for the actual photodesorption in protoplanetary disks. For example, \citet{imanishi_2003} observed that Class I YSOs might have higher X-ray luminosity distributions than Class II and Class III YSOs. \citet{hamaguchi_xray_2005} suggested that Herbig Ae/Be stars could emit X-rays with higher luminosity than low-mass pre-main sequence stars such as T Tauri stars. 
\\\\
This method outputs values of the same order of magnitude for the different X-ray emission spectra. However, we should note that this computation has some limitations:
\\\\
- the attenuation factor of gas and dust applied to the X-ray emission spectra could vary from disk to disk because of differences in gas and dust densities and disk geometry, as has been shown in protoplanetary disk modeling (e.g., \citealt{agundez_chemistry_2018, walsh_molecular_2015}).
\\\\
- the photodesorption yield at 564 eV was derived under our experimental conditions using an X-ray flux of $\sim 10^{13}$ photon/s. According to what we discussed in the previous sections, using a higher flux (e.g., higher by approximately two orders of magnitude) in similar experiments \citep{ciaravella_x-ray_2020} outputs different and important results: the X-ray photodesorption of methanol, formic acid, dimethyl ether, and/or ethanol, for example, appears to be more easily detected in our experiments, with a lower X-ray flux. In protoplanetary disk environments, the local X-ray flux that could be found in the cold regions (T < 30 K), which is between $10^{-6}$ and $10^{-2}$ erg/cm$^2$/s (i.e., between $10^{3}$ and $10^{7}$ photon/cm$^2$/s) in the case of a T Tauri-like disk modeling \citep{walsh_chemical_2012}, for instance, is lower by many orders of magnitude than the flux used under experimental conditions. As irradiating molecular ices with a high X-ray flux during a short time (experimental conditions) may result in significantly different outputs than irradiating these ices with a lower flux but during a much longer time and with other physical processes at play (astrophysical conditions), the experimental results should be extrapolated to astrophysical environments with care. 
\\\\
In the previous sections, we have demonstrated that X-rays are more efficient than UV photons in photodesorbing COMs (methanol, formic acid, dimethyl ether, and/or ethanol) from pure methanol ice per absorbed photon (this is also the case per incident photon) and under experimental conditions. When these experimental results are extrapolated to protoplanetary disk environments, regarding the estimated astrophysical UV photodesorption yield of methanol from pure methanol ice in \citet{bertin_uv_2016}, which is $1.5 \times 10^{-5}$ molecule/photon, our results in Table \ref{Astro_Yields} also suggest that given an incident photon, X-rays are at least as efficient as UV photons in photodesorbing methanol from pure methanol ice. However, as these yields are computed per incident photon, we should also consider the differences between the local UV flux and the local X-ray flux in protoplanetary disks. In the T Tauri protoplanetary disk model implemented by \citet{walsh_chemical_2012}, the cold regions (T < 50 K) near the disk midplane lie at a radius greater than 10 AU. For $Z/R$ < 0.2, X-rays emitted from the young star reach this region with a flux between 10$^{-6}$ and 10$^{-4}$ erg.cm$^{-2}$.s$^{-1}$ (10$^{3}$ and 10$^{5}$ photon.cm$^{-2}$.s$^{-1}$), and the UV flux is negligible (< 10$^{2}$ photon.cm$^{-2}$.s$^{-1}$). According to our results, X-ray photodesorption from the typical interstellar ices that could be found in this region should participate in the richness and diversity of the gas phase in addition to possible cosmic ray-induced sputtering \citep{dartois_cosmic_2018, dartois_non-thermal_2019}. In the outer cold regions, away from the midplane (for $Z/R$ > 0.2), the UV flux is non-negligible and comparable with the X-ray flux (from 10$^{4}$ to 10$^{7}$ photon.cm$^{-2}$.s$^{-1}$). In these regions, our estimated astrophysical photodesorption yields also indicate that X-ray photodesorption per incident photon is as efficient as UV photodesorption. In addition to the stellar X-ray emission, \citet{rab_x-ray_2018} showed using a T Tauri-like protoplanetary disk model and results from \citet{adams_background_2012} that the X-ray background field (XBGF) possibly originating from clusters of YSOs surrounding the protoplanetary disk could dominate in terms of flux the central young star X-ray emission at the very outer cold regions of the disk (at a radius > 10 AU, for any $Z/R$). This background field could also act as an efficient source for X-ray photodesorption. \\

\begin{table}[!h]
\caption{Average astrophysical photodesorption yield in molecule/photon extrapolated from our experimental results using the method described in section 2 for different molecules at 15 K. The X-ray emission spectrum used is the TW Hya one from \citet{nomura_molecular_2007}, to which we applied an attenuation factor corresponding to $n_H = 10^{23}$ cm$^2$.}

\begin{center}
\begin{threeparttable}

\begin{tabular}{p{2cm}p{2,5cm}p{3cm}}

\hline \hline \\
\centering Ice & \centering Desorbed Molecule & X-ray astrophysical desorption yield   
\\[0,2cm]\hline\\

\multirow{6}{*}{Pure methanol} & \centering H$_2$O  &  2.3 $\pm$ 1.2 $\times 10^{-4}$  \\[0,1cm] 

& \centering CO  &  3.9 $\pm$ 2.0 $\times 10^{-3}$ \\[0,1cm]
 
& \centering CO$_2$  &  7.4 $\pm$ 3.7 $\times 10^{-4}$ \\[0,1cm]
  
& \centering H$_2$CO  &  3.6 $\pm$ 1.8 $\times 10^{-5}$  \\[0,1cm]

& \centering CH$_3$OH  &  2.4 $\pm$ 1.2 $\times 10^{-4}$   \\[0,1cm]

& \centering HCOOH, C$_2$H$_6$O  & 2.4 $\pm$ 1.2 $\times 10^{-5}$  \\[0,1cm]

\hline
          
\end{tabular}
   
\end{threeparttable}
\end{center}   
\label{Astro_Yields_I}

\end{table}

In Table \ref{Astro_Yields_I} we display the astrophysical photodesorption yields corresponding to the main molecules observed in our experiments. These yields are derived with the same method as for Table \ref{Astro_Yields}. As we do not see any significant effect of the X-ray emission spectrum used on the estimated yields in Table \ref{Astro_Yields}, we decided to use the X-ray emission spectrum of the TW Hya star \citep{nomura_molecular_2007} alone, with an attenuation factor corresponding to $n_H = 10^{23} cm^2$. The yields corresponding to other attenuation factors can be deduced by applying the same variations as in Table \ref{Astro_Yields}. These yields, however, should be considered with caution because they are extracted from pure methanol ices that are unlikely to be found in the interstellar medium. More realistic systems, in which methanol would be embedded in a frozen matrix composed of the main species of the interstellar ices (such as H$_2$O, CO, or CO$_2$), may exhibit photodesorption efficiency different from pure methanol ice. We established in section 4 that the photodesorption of neutrals in the X-ray range is much likely carried out by the thermalization of high-energy electrons into the molecular solids, and involves subsequent chemistry, as is demonstrated by the desorption of more complex molecules than CH$_3$OH. It is therefore expected that the composition of the ice affects the photodesorption efficiency. In order to access more relevant photodesorption yields, it is then necessary to constrain this composition effect by studying model binary ices containing methanol. This has been achieved, and the outcomes are presented in a second article (paper II).

\section{Conclusion}
Pure methanol ice was irradiated by monochromatic X-rays in the range of 525-570 eV. Intact methanol, other COMs, and simpler molecules were found to photodesorb due to X-ray absorption of core O(1s) electrons, quantified by TEY measurement, which leads to a cascade of low-energy secondary electrons within the ice. X-ray photodesorption yields were derived and found to be intimately linked to X-ray induced chemistry, which indicates that X-ray induced electron-stimulated desorption (XESD) may be the dominant mechanism in X-ray photodesorption from these ices. However, electron-stimulated desorption experiments are mandatory to conclude on the dominant mechanism. The main conclusions of this paper are listed below.

\begin{enumerate}

\item CH$_3$OH X-ray photodesorption from pure methanol ice is found to be efficient with a yield of $\sim 10^{-2}$ molecule desorbed by incident photon and is assumed to be due to the recombination of CH$_2$OH and/or CH$_3$O into CH$_3$OH.

\item X-ray photodesorption of formic acid, ethanol, and/or dimethyl ether is detected with a yield of $\sim 10^{-3}$ molecule desorbed per incident photon.

\item Destruction of methanol molecules by photolysis and/or radiolysis seems to defavor the detection of methanol desorption in X-ray irradiation experiments, and the fluence received by the ice or the X-ray flux used under experimental conditions may be a critical parameter for detecting methanol desorption. 

\item X-ray photodesorption is found to be more efficient by more than one order of magnitude than UV photodesorption \citep{bertin_uv_2016} for pure methanol ice per incident or absorbed photon, which is assumed to be due to a difference in the induced chemistry. When these experimental results were extrapolated to a protoplanetary disk environment, we found that X-rays are at least as efficient as UV photons in desorbing molecules from interstellar ices depending on the region considered. 
\end{enumerate}
The yields we presented, extracted from pure methanol ices, should be considered with caution because the photodesorption efficiency may depend on the composition of the more complex interstellar ices containing methanol. This is discussed in paper II.

\begin{acknowledgements}
This work was done with financial support from the Region Ile-de-France
DIM-ACAV+ program and by the European Organization for Nuclear Research
(CERN) under the collaboration Agreement No. KE3324/TE. We would like to
acknowledge SOLEIL for provision of synchrotron radiation facilities
under Project Nos. 20181140, and we thank N. Jaouen, H. Popescu and R.
Gaudemer for their help on the SEXTANTS beam line. This work was
supported by the Programme National “Physique et Chimie du Milieu
Interstellaire” (PCMI) of CNRS/INSU with INC/INP co-funded by CEA and
CNES. Financial support from the LabEx MiChem, part of the French state
funds managed by the ANR within the investissements d avenir program
under Reference No. ANR-11-10EX-0004-02, is gratefully acknowledged.
\end{acknowledgements}

\bibliographystyle{aa}
\bibliography{Bibli}

\end{document}